\documentclass[]{article}
\usepackage{amsmath}
\usepackage{amssymb}
\usepackage{authblk}
\usepackage{caption}
\usepackage{cite}
\usepackage{xcolor}
\usepackage{color}
\usepackage{float}
\usepackage{geometry}
\usepackage{graphicx}
\usepackage{hyperref}
\usepackage{textcomp}
\usepackage{epstopdf}
\usepackage{subcaption}
\newcommand\hl[1]{%
	\bgroup
	\hskip0pt\color{blue!80!black}%
	#1%
	\egroup
}
\title{Solutions of a charged scalar field in five-dimensional helicoid solution with electromagnetic field}
\author{Tolga Birkandan\thanks{E-mail: birkandant@itu.edu.tr, ORCID: 0000-0003-4434-2259.} \\
	Department of Physics, Istanbul Technical University, 34469 Istanbul,
	Turkey.}

\begin{document}
\maketitle
\text{Keywords: Klein-Gordon equation, gravitational instanton, exact solutions, numerical solutions}
\begin{abstract}
\noindent We study a charged and massive scalar field in the background of the Nutku-Ghezelbash-Kumar metric which is obtained by the addition of a time coordinate to the Nutku helicoid metric in a non-trivial way. The angular part of the Klein-Gordon equation can be written as a double confluent Heun equation. The radial equation cannot be solved in terms of a known function in its general form. However, in some special cases, the radial equation can also be written explicitly as a double confluent Heun equation. We study the full radial equation numerically and observe that the electromagnetic field parameter defines an effective cut-off on the range of the radial coordinate. Finally, we obtain a quasi-exact solution with an approximation.
\end{abstract}
%\begin{description}
%\item[Keywords: Scalar field, instanton metric, Klein-Gordon equation, Einstein-Maxwell equation] 
%\end{description}
%%%%%%%%%%%%%%%%%%%%%%%%%%%%%%%%%%%%%%%%%%%%%%%%%%%%%%%%%%%%%%%%%%%%%%%%%%%%%%%%%%%%%
\section{Introduction}
Gravitational instantons are Euclidean exact solutions of Einstein's field equations in vacuum which admit hyper-K{\"a}hler structure and their properties can be studied by a Euclidean version of the Newman-Penrose formalism \cite{Eguchi:1980jx,Aliev:1998cu,Goldblatt:1994rx,Goldblatt:1994iw,Birkandan:2007cx}.

Nutku's instanton metric corresponds to the helicoid minimal surface which is a realization of Weierstrass’ general local solution for minimal surfaces. It has a quadratic Killing tensor and a curvature singularity at the origin. The metric corresponds to the catenoid minimal surface with a reparametrization \cite{Nutku:1996dn,Aliev:1996sf}. The symmetries and the scalar Green's function corresponding to this metric are studied in \cite{Aliev:1998xu}.

Unfortunately, the literature on the Nutku solution is limited. The first group of studies focuses on the solutions of the wave equations in this background. Klein-Gordon and Dirac equations have been studied for the Nutku helicoid metric in its original form, and they have been solved in terms of Mathieu functions \cite{Aliev:1996sf,Sucu:2004qx,Birkandan:2006ac}. A time coordinate can be added to the metric in a trivial way as $ds^2=-dt^2+ds_{\text{Nutku}}^2$ and the wave equations can be solved in terms of the double confluent Heun (DCH) functions or the Mathieu functions as in the original case \cite{Birkandan:2006ac,Birkandan:2007ey,Birkandan:2007cw}. The second group of studies involves the Nutku solution as a base for higher-dimensional metrics. Ghezelbash and Kumar constructed new higher-dimensional metrics based on the Nutku solution in Einstein-Maxwell theory \cite{Ghezelbash:2017bjs}, Butler and Ghezelbash employed the Nutku metric in constructing new solutions to five-dimensional generalized Einstein-Maxwell-dilaton theory with cosmological constant and two coupling constants \cite{Butler:2018xdo}, and Ghezelbash embedded the Nutku solution in the membranes and 5-branes of the eleven-dimensional supergravity \cite{Ghezelbash:2022tbb}.

In this paper, we study a charged and massive scalar field in the background of the Nutku-Ghezelbash-Kumar (NGK) metric which is based on the Nutku helicoid metric with a time coordinate added in a non-trivial manner in the Einstein-Maxwell theory \cite{Ghezelbash:2017bjs}. The aforementioned works of the Ghezelbash group have added new Nutku-based metrics to the literature and our analysis will be the first step in studying the wave equations associated with them. Our study is based on studying the types of solutions of the Klein-Gordon equation in this new background. NGK metric is a solution in the Einstein-Maxwell theory which enables us to observe the properties of a charged scalar field in a Nutku-based metric for the first time in the literature. We aim to observe the effects of the non-trivially added time coordinate and the electromagnetic potential on the solutions of the Klein-Gordon equation and compare the results with the literature.

The angular part of the Klein-Gordon equation studied in this background yields a DCH equation. In some limits, the radial part can be written as a DCH equation, and the numerical solution of the radial equation also signals a DCH-like behavior. Thus, we observe that the DCH-type behavior of the wave solutions persists in the non-trivial Nutku-based Einstein-Maxwell theory given by the five-dimensional NGK metric.

Heun-type equations are known to appear in many applications in gravity, especially as solutions of the wave equations \cite{Ronveaux,Slavyanov,Hortacsuheun,Birkandan:2017rdp,Batic:2007it,Dariescu:2018rrr,Dariescu:2021zve}. The emergence of the hypergeometric functions as solutions of the wave equations can be a sign of the conformal symmetry \cite{Birkandan:2015yda}, whereas Heun-type functions are not associated with such symmetries, yet. Besides, all studies based on the Nutku helicoid metric in the literature share DCH functions as solutions of the wave equations. Our results show that in a non-trivial Einstein-Maxwell theory based case, the exact solutions of a charged scalar field are still DCH functions for the angular part of the wave equation and for the limiting cases of the radial part. In the presence of the electromagnetic interaction, the DCH-like behavior is not exact and the radial equation cannot be solved in terms of DCH functions. Thus we can interpret our result with the electromagnetic interaction as the first deviation from the exact DCH behavior in the works based on the Nutku helicoid metric. Our motivation for studying a charged scalar field in the background of the NGK spacetime is limited with the types of the solutions and we manage to find the first deviation from the exact DCH behavior in the literature.

The paper is organized in the following way: In the second section, we present the five-dimensional NGK metric, in the third section we study the charged and massive scalar field by separating the Klein-Gordon equation into its angular and radial parts where we obtain exact and quasi-exact solutions for some limits. In this section, we also study the full radial equation numerically to observe an effective radial cut-off. We summarize our results in the conclusions section, and we present the transformation of the DCH equation to the Mathieu equation for a special case that we observe in the studies associated with the Nutku instanton in an appendix.
%%%%%%%%%%%%%%%%%%%%%%%%%%%%%%%%%%%%%%%%%%%%%%%%%%%%%%%%%%%%%%%%%%%%%%%%%%%%%%%%%%%%%
\section{The 5D helicoid metric with EM field}
The Nutku helicoid solution is obtained from a general instanton metric 
\begin{align}
ds^2=\frac{1}{P} \bigg\{\bigg[ \bigg( \frac{\partial f}{\partial \tau}  \bigg)^2 +\kappa \bigg] (d\tau^2+dy^2) +\bigg[1+\bigg( \frac{\partial f}{\partial x}  \bigg)^2 \bigg](dx^2+dz^2)+2\bigg( \frac{\partial f}{\partial \tau}  \bigg)\bigg( \frac{\partial f}{\partial x}  \bigg)(d\tau dx+dydz)\bigg\},
\end{align}
where $f=f(\tau,x)$ is the Monge ansatz which defines the surface, and $P=P(\tau,x)=\sqrt{1+\kappa (\frac{\partial f}{\partial \tau})^2+(\frac{\partial f}{\partial x})^2}$. Einstein field equations reduce to
\begin{equation}
\bigg[1+\bigg(\frac{\partial f}{\partial x}\bigg)^2\bigg]\bigg(\frac{\partial^2 f}{\partial \tau^2} \bigg)-2\bigg(\frac{\partial f}{\partial \tau}\bigg)\bigg(\frac{\partial f}{\partial x}\bigg)\bigg(\frac{\partial^2 f}{\partial \tau \partial x} \bigg)+\bigg[ \bigg( \frac{\partial f}{\partial \tau}  \bigg)^2 +\kappa \bigg]\bigg(\frac{\partial^2 f}{\partial x^2} \bigg)=0.
\end{equation}
One takes $\kappa=+1$ for minimal surfaces in $\mathbb{R}^3$ and $\kappa=-1$ corresponds to Born-Infeld equation. The helicoid minimal surface is governed by $\kappa=+1$ and $f(\tau,x)=a[tan^{-1}(\frac{x}{\tau})]$. By setting $x=rcos(\theta)$ and $\tau=rsin(\theta)$, we get
\begin{align}\label{metrikkk}
	ds^2=\frac{1}{\sqrt{1+\frac{a^2}{r^2}}}\bigg[dr^2+(a^2+r^2)d\theta^2+\bigg(1+\frac{a^2sin^2\theta}{r^2} \bigg)dy^2+ \bigg(1+\frac{a^2cos^2\theta}{r^2} \bigg)dz^2-\frac{a^2sin(2\theta)dydz}{r^2}\bigg],
\end{align}
which is the Nutku helicoid solution. The definition ranges of the coordinates will be specified shortly. The parameter $a$ in the metric (\ref{metrikkk}) is associated with the helicoidal geometry of the space. A nonzero $a$ parameter ensures the helicoidal (or catenoidal) surface and $a=0$ corresponds to a plane. If we replace $a^2$ with $-a^2$ in the metric, we obtain the catenoid case \cite{Nutku:1996dn,Aliev:1998xu,Ghezelbash:2017bjs}.

The Nutku helicoid metric with a non-trivial addition of the time coordinate (the NGK metric) is given as
\begin{align}\label{metrikk}
ds^2=&-\frac{dt^2}{(H(r))^2}+\frac{H(r)}{\sqrt{1+\frac{a^2}{r^2}}}\bigg[dr^2+(a^2+r^2)d\theta^2 \nonumber \\ 
&+\bigg(1+\frac{a^2sin^2\theta}{r^2} \bigg)dy^2+ \bigg(1+\frac{a^2cos^2\theta}{r^2} \bigg)dz^2-\frac{a^2sin(2\theta)dydz}{r^2}\bigg],
\end{align}
where $r \in (0,\infty)$, $\theta \in [0,2\pi]$, and the Killing coordinates $y$ and $z$ are periodic on a 2-torus \cite{Ghezelbash:2017bjs}.
The NGK metric is a solution of Einstein-Maxwell equations. For the helicoid case, the metric function
\begin{equation}
H(r)=1+a c_2 sinh^{-1} \bigg( \frac{r}{a} \bigg),
\end{equation}
is related with the electromagnetic (EM) potential. The only non-zero component of the EM potential is
\begin{equation}
A_t=\sqrt{\frac{3}{2}} \frac{1}{H(r)},
\end{equation}
in five-dimensions as given in \cite{Ghezelbash:2017bjs}. In order to prevent a signature change in the metric, $a \geq 0$ and $c_2 \geq 0$ will be required. 

The Kretschmann scalar yields a lengthy numerator, whereas the denominator of the Kretschmann scalar reads $4r^6(a^2+r^2)^3H(r)^6$. Then the Kretschmann scalar of the NGK solution shares the curvature singularity of the original Nutku metric at $r=0$. However, we must emphasize that the point $r=0$ is not included in the radial range of the NGK metric which is given by $0<r<\infty$ in \cite{Ghezelbash:2017bjs}. If the point $r=0$ were in the radial range, then it would be a naked singularity as the metric functions with $a \geq 0$ do not yield a non-imaginary event horizon that shields this singularity. In this present form, the NGK metric is a five-dimensional solution of the Einstein-Maxwell theory based on the helicoid geometry with no black hole interpretation. As noted in \cite{Ghezelbash:2017bjs}, a submanifold of the five-dimensional helicoidal solution is conformally equivalent to a wormhole handle $[dr^2+(a^2+r^2)d\theta^2]$ \cite{Harris:1993ds,Ellis:1973yv}, and this can be regarded as a link to a more conventional structure.

\begin{figure}[H]
	\centering
	\includegraphics[scale=0.6]{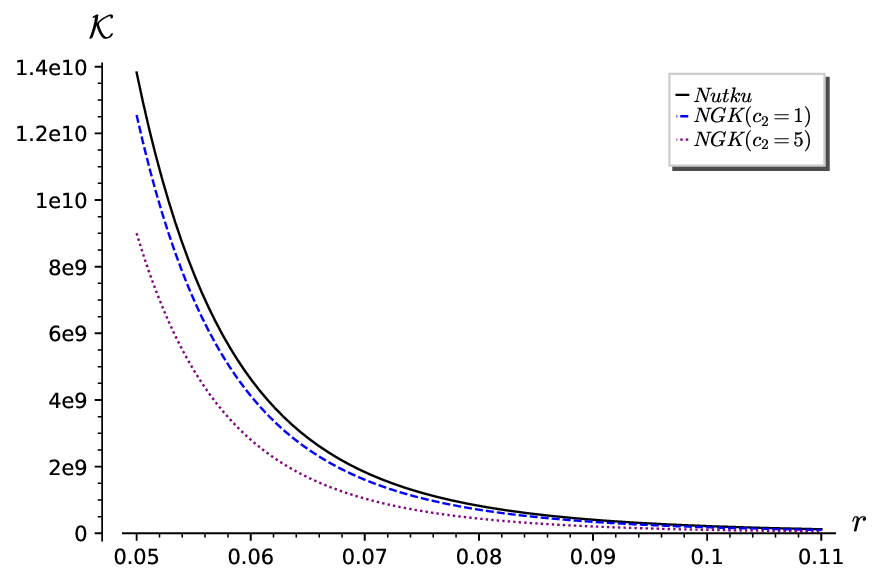} 
	\caption{Kretschmann scalar $\mathcal{K}$ \{$a=3$\}.}
	\label{fig:kscalarfig}
\end{figure}

A plot of the Kretschmann scalar for the Nutku helicoid solution and the NGK metric can be seen in Figure (\ref{fig:kscalarfig}).
The original Nutku helicoid metric is asymptotically Euclidean \cite{Nutku:1996dn,Aliev:1998xu,Ghezelbash:2017bjs}, and the limit $a \rightarrow 0$ yields the Minkowski spacetime for the NGK metric \cite{Ghezelbash:2017bjs}.
%%%%%%%%%%%%%%%%%%%%%%%%%%%%%%%%%%%%%%%%%%%%%%%%%%%%%%%%%%%%%%%%%%%%%%%%%%%%%%%%%%%%%
\section{The charged scalar field}
The Klein-Gordon equation for a massive and charged scalar field $\Phi$ can be given as
\begin{equation}
[(\nabla ^\nu-iqA^\nu)(\nabla_\nu-iqA_\nu)-\mu^2]\Phi(t,r,\theta,y,z)=0,
\end{equation}
where $\mu$ is the mass and $q$ is the charge of the scalar field. 

The equation reads
\begin{eqnarray}\label{fullpde}
&&\bigg\{-(H(r))^2 \frac{\partial ^2}{\partial t^2}+i\sqrt{6}qH(r) \frac{\partial}{\partial t}+ \frac{\sqrt{a^2+r^2}}{rH(r)} \frac{\partial ^2}{\partial r^2}+\frac{1}{\sqrt{a^2+r^2}H(r)}  \frac{\partial}{\partial r} \nonumber\\
&&+\frac{1}{\sqrt{a^2+r^2}rH(r)} \bigg[ \frac{\partial ^2}{\partial \theta^2}
+(a^2 cos^2(\theta)+r^2) \frac{\partial ^2}{\partial y^2}+(a^2 sin^2(\theta)+r^2) \frac{\partial ^2}{\partial z^2}+(2a^2 sin(\theta)cos(\theta))\frac{\partial ^2}{\partial y \partial z} \bigg] \nonumber\\
&&+\bigg( \frac{3q^2}{2}-\mu^2 \bigg) \bigg\} \Phi(t,r,\theta,y,z) =0.
\end{eqnarray}
One can directly solve this partial differential equation numerically to study the behavior of the scalar field in the background of the NGK metric. However, we need to study the Klein-Gordon equation by separating it into radial and angular parts in order to see if it has exact or quasi-exact solutions and whether the Heun-type solutions that have emerged in the previous works persist in the NGK metric case. Following Section 5 of Reference \cite{Aliev:1998xu}, for the Nutku helicoid metric, we see that the existence of a Killing tensor which is related with the angular part of the Klein-Gordon equation and the vector fields which correspond to the directions $y$ and $z$ lead the separation of the scalar field. Comparing the results given in \cite{Aliev:1998xu} with our case, we observe that the Killing tensor written for the Nutku helicoid case, namely, $\hat{K}=-[\frac{\partial ^2}{\partial \theta^2}+a^2 cos^2(\theta) \frac{\partial ^2}{\partial y^2}+a^2 sin^2(\theta) \frac{\partial ^2}{\partial z^2}+2a^2 sin(\theta)cos(\theta)\frac{\partial ^2}{\partial y \partial z}]$, is still applicable in the NGK metric as well as the vector fields $\frac{\partial}{\partial y}$ and $\frac{\partial}{\partial z}$ with an additional time-related vector field $\frac{\partial}{\partial t}$. Thus, we can use the ansatz,
\begin{equation}\label{ansatz}
\Phi(t,r,\theta,y,z)=e^{-i \omega t} e^{i m_1 y} e^{i m_2 z} \varphi(r,\theta),
\end{equation}
which yields
\begin{align}
\bigg\{& (a^2+r^2)\frac{\partial ^2}{\partial r^2} + \frac{\partial^2}{\partial\theta^2}+r\frac{\partial}{\partial r}+r\sqrt{a^2+r^2}H(r) \bigg(\sqrt{6}qH(r)\omega+(H(r))^2\omega^2+\frac{3}{2}q^2-\mu^2 \bigg)-(m_1^2+m_2^2)r^2 \nonumber\\
&-a^2(m_1^2-m_2^2)cos^2(\theta)-2a^2m_1m_2sin(\theta)cos(\theta)-a^2m_2^2  \bigg\}\varphi(r,\theta)=0.
\end{align}
The radial and angular parts can be separated using $\varphi(r,\theta)=R(r)S(\theta)$, namely,
\begin{align}
&\frac{d^2R(r)}{dr^2}+\frac{r}{a^2+r^2}\frac{dR(r)}{dr}+\frac{r\sqrt{a^2+r^2}H(r) \bigg[\sqrt{6}qH(r)\omega+(H(r))^2\omega^2+\frac{3}{2}q^2-\mu^2 \bigg]-(m_1^2+m_2^2)r^2-\lambda}{a^2+r^2} R(r)=0, \label{radial}\\
&\frac{d^2 S(\theta)}{d\theta^2}-\bigg[a^2(m_1 cos \theta+m_2 sin \theta)^2 -\lambda\bigg]S(\theta)=0, \label{angulareqn}
\end{align}
where $\lambda$ is the separation constant.
%%%%%%%%%%%%%%%%%%%%%%%%%%%%%%%%%%%%%%%%%%%%%%%%%%%%%%%%%%%%%%%%%%%%%%%%%%%%%%%%%%%%%
\subsection{The angular equation}
Before attempting to solve its general form, let us study some special cases and the numerical behavior of the angular equation given in (\ref{angulareqn}). 

The general form of the Mathieu equation is \citenum{nist}
\begin{equation}\label{mathieum}
\frac{d^2S}{d\theta^2}+[b-2qcos(2\theta)]S=0,
\end{equation}
and it can be solved as
\begin{equation}
S(\theta)=C_1 Se(b,q,\theta)+C_2 So(b,q,\theta),
\end{equation}
where $Se$ and $So$ are the even and odd Mathieu functions, respectively. Here, $C_1$ and $C_2$ are constants and the functions satisfy $Se(b,q,-\theta)=Se(b,q,\theta)$ and $So(b,q,-\theta)=-So(b,q,\theta)$. 

For $m_1=0$, the solution of (\ref{angulareqn}) reads
\begin{equation}
	S(\theta)=C_1 Se \bigg(\lambda-\frac{a^2 m_2^2}{2},-\frac{a^2 m_2^2}{4},\theta \bigg)+C_2 So \bigg(\lambda-\frac{a^2 m_2^2}{2},-\frac{a^2 m_2^2}{4},\theta \bigg),
\end{equation}
and for $m_2=0$, we have
\begin{equation}
	S(\theta)=C_1 Se \bigg(\lambda-\frac{a^2 m_1^2}{2},\frac{a^2 m_1^2}{4},\theta \bigg)+C_2 So \bigg(\lambda-\frac{a^2 m_1^2}{2},\frac{a^2 m_1^2}{4},\theta \bigg),
\end{equation}
as the solution of equation (\ref{angulareqn}).

Another special case with $m_1=m_2\neq 0$ is not as straightforward as the forms above. We can revisit this case after completing the analysis of the full angular equation. 

\begin{figure}[H]
	\begin{subfigure}{.5\textwidth}
		\centering
		% include first image
		\includegraphics[width=1.0\linewidth]{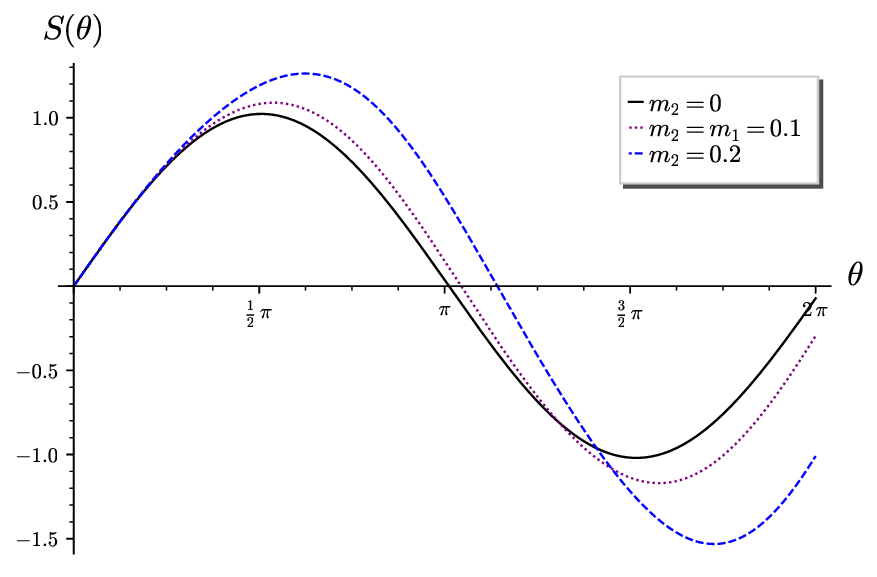}
		\caption{Varying $m_2$ ($m_1=0.1$).}
		\label{fig:sub-firstang0}
	\end{subfigure}
	\begin{subfigure}{.5\textwidth}
		\centering
		% include second image
		\includegraphics[width=1.0\linewidth]{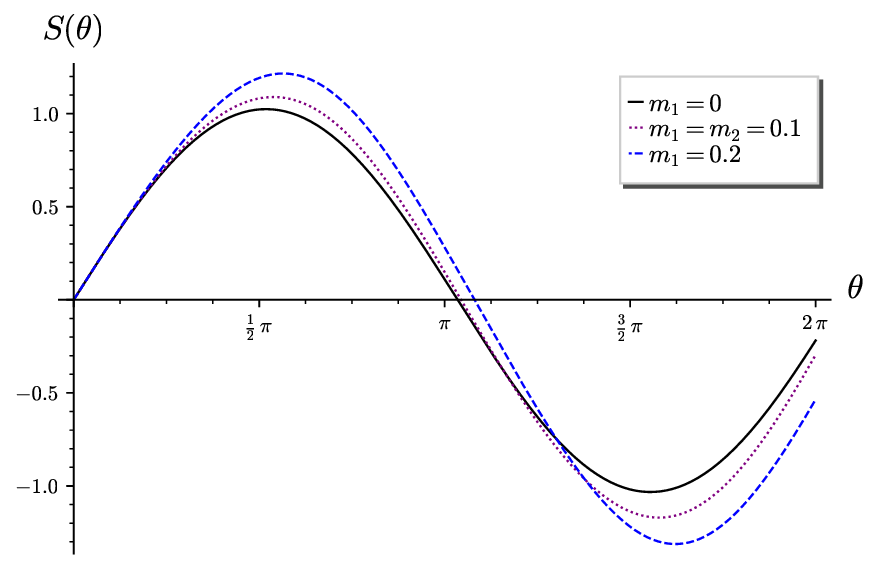}
		\caption{Varying $m_1$ ($m_2=0.1$).}
		\label{fig:sub-secondang0}
	\end{subfigure}
	\caption{Numerical solutions of the angular equation (\ref{angulareqn}) with \{$\lambda=1, a=3$\}.}
	\label{fig:varangular}
\end{figure}

One can directly solve the equation (\ref{angulareqn}) numerically to observe the general angular behavior and the effects of the parameters $m_1$ and $m_2$ on the solution. Using the SageMath computer algebra system \cite{sagemath}, we obtain Figure (\ref{fig:varangular}).
Figure (\ref{fig:varangular}) signals that all cases have similar solution types. The general analysis below will explicitly prove that all cases share the Mathieu-type solutions.

The angular part of the Klein-Gordon equation can be understood better if we perform the change of variable $v= -tanh(i \theta)$. Then we obtain
\begin{align}\label{angeqn}
\frac{d^2 S(v)}{dv^2}+\frac{2v}{v^2-1}\frac{dS(v)}{dv}+\frac{(a^2 m_2^2-\lambda) v^2-2i a^2 m_1 m_2 v-a^2 m_1^2+\lambda}{(v-1)^3 (v+1)^3} S(v)=0.
\end{align}
This equation has two irregular singular points of rank-1 located at $v = \{-1, 1\}$ which corresponds to a DCH equation \cite{Ronveaux,Slavyanov}. 

Let us define the general form of the DCH equation that has been used in studying the case where the time component was introduced in a trivial manner \cite{Birkandan:2006ac}, and this will be useful for us while comparing our solutions in the $\mu=q=c_2=0$ limit. The general form of the DCH equation reads
\begin{equation} \label{heund}
	\frac{d^2 H_D}{dx^2}
	+{\frac {2\,{x}^{5} -\alpha\,{x}^{4}-4\,{x}^{3}+2\,x+\alpha}{ \left( x-
			1 \right) ^{3} \left( x+1 \right) ^{3}}}
	\frac{dH_D}{dx}
	+{\frac {\beta{x}^{2}+ \left( \gamma+2\,\alpha \right) x+\delta}{
			\left( x-1 \right) ^{3} \left( x+1 \right) ^{3}}}
	H_D=0,
\end{equation}
and the irregular, rank-1 singular points are located at $x=\{-1, 1\}$. The radius of convergence of the local solution $H_D(\alpha,\beta,\gamma,\delta,x)$ around the ordinary point $x=0$ is $|x|<1$. For a detailed analysis of the equation, one can see \cite{Ronveaux,Slavyanov,Birkandan:2020clt}.

We can compare the equations (\ref{angeqn}) and (\ref{heund}) to find the parameters in the DCH equation, namely,
\begin{align}
\alpha&=0,\\
\beta&=a^2m_2^2-\lambda,\\
\gamma&=-2ia^2m_1m_2,\\
\delta&=\lambda-a^2m_1^2,
\end{align}
needed to show that the angular part of the Klein-Gordon equation can be solved in terms of DCH functions with these parameters.

We should note that the angular equation is related with the quadratic Killing tensor of the original Nutku metric \cite{Birkandan:2006ac,Aliev:1998xu}. It was solved in terms of Mathieu functions in \cite{Birkandan:2006ac} and here, we present it as a DCH equation. 

In \cite{Birkandan:2006ac}, it was shown that one can transform the DCH equation to a Mathieu-type equation via some transformations in the Nutku helicoid case. In our appendix, we will follow the same procedure for the general form of the DCH equation (\ref{heund}) with $\alpha=0$ to show that it can be transformed into the Mathieu equation. This transformation is also valid for our angular equation (\ref{angeqn}) by using the equation (\ref{cicimathieu}) and the future DCH-type results which will have $\alpha=0$. 

Using the DCH parameters of the angular equation in (\ref{cicimathieu}), we can observe that the case with $m_1=m_2\neq 0$ is not a special case in terms of analysis of the equation.

For $q=0$, the equation (\ref{mathieum}) has the special solutions $Se(b,0,\theta)=cos(\sqrt{b} \theta)$ and $So(b,0,\theta)=sin(\sqrt{b} \theta)$. We can analyze this limit using the DCH parameters in (\ref{cicimathieu}) to have,
\begin{equation}
q=-\frac{1}{4}\sqrt{a^4(m_1^2+m_2^2)^2}.
\end{equation}
The parameter $a$ ensures the helicoidal geometry, therefore it cannot be zero. Either $m_1$ or $m_2$ should be imaginary to satisfy the $q=0$ limit. In order to keep the oscillatory behavior of the coordinates $y$ and $z$ in the scalar solution ansatz (\ref{ansatz}), we conclude that the solutions $cos(\sqrt{b} \theta)$ and $sin(\sqrt{b} \theta)$ cannot exist for our angular equation.

\begin{figure}[H]
	\begin{subfigure}{.5\textwidth}
		\centering
		% include first image
		\includegraphics[width=1.0\linewidth]{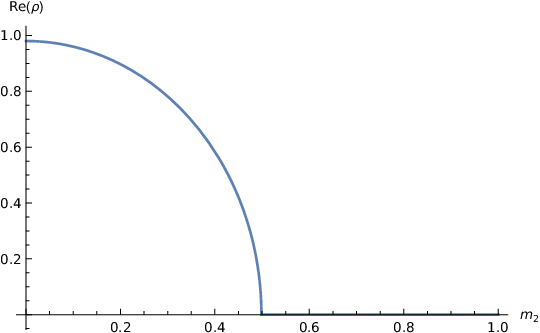}
		\caption{Real part of the characteristic exponent $\rho$.}
		\label{fig:sub-firstang00}
	\end{subfigure}
	\begin{subfigure}{.5\textwidth}
		\centering
		% include second image
		\includegraphics[width=1.0\linewidth]{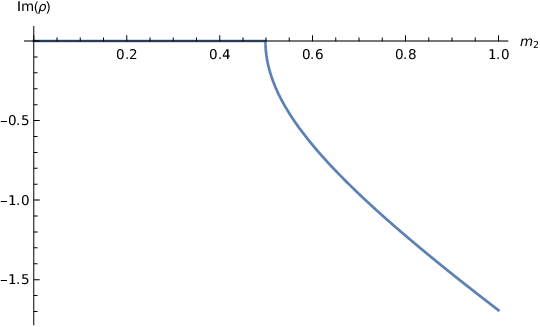}
		\caption{Imaginary part of the characteristic exponent $\rho$.}
		\label{fig:sub-secondang00}
	\end{subfigure}
	\caption{Characteristic exponent $\rho$ for varying $m_2$ with \{$\lambda=1, a=3, m_1=0.1 $\}.}
	\label{fig:char}
\end{figure}

The periodicity of the angular solutions can be analyzed using Floquet's theorem \cite{arfken}. According to this theorem, we can write the solution of the Mathieu equation (\ref{mathieum}) as $S(\theta)=e^{i\rho\theta}\psi(\theta)$ where $\psi(\theta)$ is periodic with period $\pi$ and $\rho$ is called the characteristic exponent of $S(\theta)$. Wolfram Cloud is able to calculate characteristic exponents of the Mathieu solutions \cite{wolfram}, and we plot Figure (\ref{fig:char}) using this system.
For the numerical parameters used in Figure (\ref{fig:char}), the characteristic exponent turns pure imaginary after $m_2\approx0.5$. According to Floquet's theorem, for negative imaginary $\rho$ values, the solution $S(\theta)$ grows as $\theta$ increases whereas it remains in the same magnitude when $\rho$ is real. If $\rho$ is an even or odd integer, then the period of $S(\theta)$ is $\pi$ or $2\pi$, respectively. If we have $\rho=\frac{2r}{s}$ where $r$ and $s>2$ are integers with no common divisor, then the period of the solution $S(\theta)$ is $s\pi$ which corresponds to our case. 

\begin{figure}[H]
	\begin{subfigure}{.5\textwidth}
		\centering
		% include first image
		\includegraphics[width=1.0\linewidth]{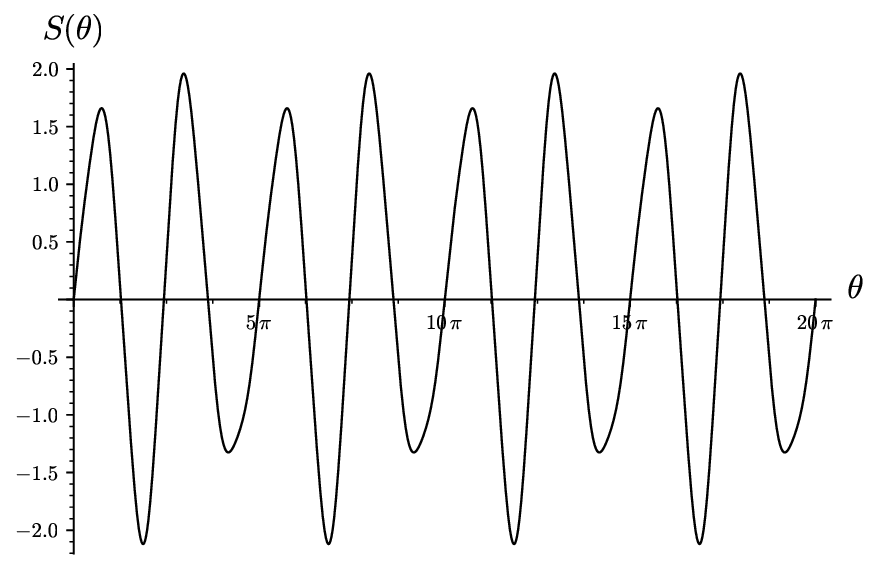}
		\caption{Angular solution with  \{Re$(\rho) \approx\ 0.8, m_2=0.287505\}$.}
		\label{fig:sub-firstang000}
	\end{subfigure}
	\begin{subfigure}{.5\textwidth}
		\centering
		% include second image
		\includegraphics[width=1.0\linewidth]{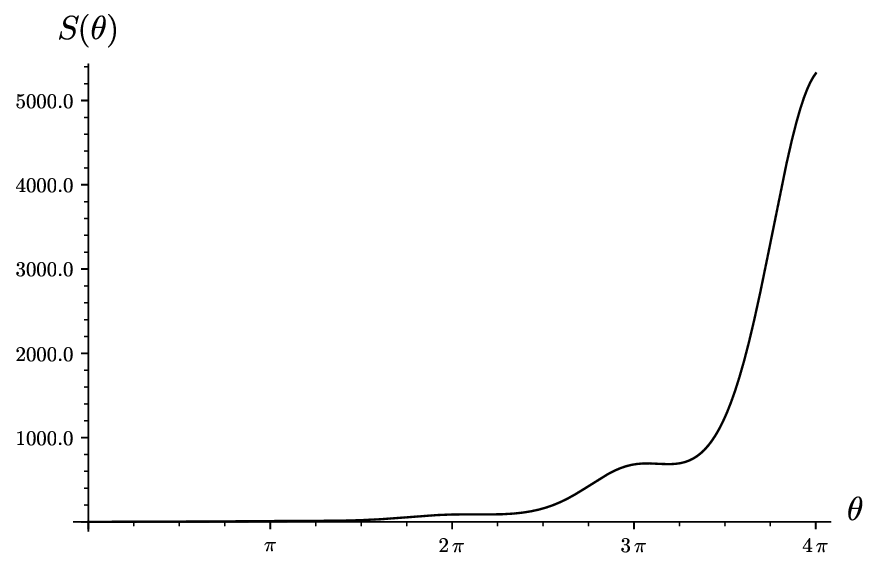}
		\caption{Angular solution with  \{Im$(\rho) \approx\ -0.85, m_2=0.6\}$.}
		\label{fig:sub-secondang000}
	\end{subfigure}
	\caption{Angular behavior and periodicity with \{$\lambda=1, a=3, m_1=0.1 $\}.}
	\label{fig:periodicity}
\end{figure}

In Figure (\ref{fig:sub-firstang000}), we observe the periodicity of the angular solution with period $5\pi$ as Re$(\rho) \approx\ \frac{4}{5}=0.8$ (i.e. $r=2, s=5$) for $m_2=0.287505$. Figure (\ref{fig:sub-secondang000}) is an example of increasing solution with $\theta$ for $m_2=0.6$ and Im$(\rho) \approx\ -0.85$.
%%%%%%%%%%%%%%%%%%%%%%%%%%%%%%%%%%%%%%%%%%%%%%%%%%%%%%%%%%%%%%%%%%%%%%%%%%%%%%%%%%%%%
\subsection{The radial equation}
If we perform the change of variable $r = asinh(x)$ in Eq. (\ref{radial}), we can write the EM metric parameter $H(r)$ as
\begin{equation}
H(x)=1+c_2 a x.
\end{equation}
and we get
\begin{equation}\label{yardimci}
\frac{d^2 R(x)}{dx^2}+\mathcal{F}(x)R(x)=0,
\end{equation}
where
\begin{align}
\mathcal{F}(x)=&\frac{a^2}{2}(m_1^2+m_2^2)-\lambda-\frac{a^2}{2}(m_1^2+m_2^2)cosh(2x) +\frac{a^2}{2} \bigg( \sqrt{6} q \omega (H(x))^2 +  \omega^2 (H(x))^2  +\frac{3}{2}q^2 - \mu^2  \bigg) sinh(2x).
\end{align}
Another change of coordinate $x = tanh^{-1}(u)$, brings the radial equation (\ref{yardimci}) in a form that the singularities can be analyzed better. The new form of the radial equation can be given as
\begin{align}\label{inceleradial}
\frac{d^2 R(u)}{du^2}+\frac{2u}{u^2-1}\frac{dR(u)}{du}+\frac{F_1(u)+F_2(u,c_2)}{(u-1)^3 (u+1)^3}R(u)=0,
\end{align}
where
\begin{align}
&F_1(u)=[(m_1^2+m_2^2)a^2-\lambda] u^2 - a^2 \bigg(\frac{3}{2} q^2-\sqrt{6} q \omega +\omega^2 - \mu^2 \bigg) u+\lambda,\\
&F_2(u,c_2)=\bigg[-a^2 \omega^2 (tanh^{-1}(u))^2  c_2^2 - a \omega (\sqrt{6} q+3 \omega) tanh^{-1}(u)   c_2-(2 \sqrt{6} q \omega-\mu^2+\frac{3}{2} q^2+3\omega^2)  \bigg]a^3 c_2 u tanh^{-1}(u). \label{F2uc2}
\end{align}
The equation (\ref{inceleradial}) has two irregular singular points of rank-1 located at $u = \{-1, 1\}$, which is the signature of a DCH equation. The range of the radial coordinate is transformed as $u \in (0,1)$.

If we set $c_2=0, q=0, \mu=0$, then we get
\begin{align}
	\frac{d^2 R(u)}{du^2}+\frac{2u}{u^2-1}\frac{dR(u)}{du}+\frac{[(m_1^2+m_2^2)a^2-\lambda] u^2 - a^2 \omega^2 u+\lambda}{(u-1)^3 (u+1)^3}R(u)=0,
\end{align}
which corresponds to the following set of the DCH parameters in Eq. (\ref{heund}),
\begin{align}
\alpha&=0,\\
\beta&=(m_1^2+m_2^2)a^2-\lambda,\\
\gamma&=- a^2 \omega^2,\\
\delta&=\lambda.
\end{align}
This result agrees with the result of the case that the time coordinate has been added in a trivial manner in \cite{Birkandan:2006ac}.

The case with $c_2=0, q \ne 0, \mu \ne 0$ also yields the DCH equation with parameters
\begin{align}
\alpha&=0,\\
\beta&=(m_1^2+m_2^2)a^2-\lambda,\\
\gamma&=- a^2 \bigg(\frac{3}{2} q^2+\sqrt{6} q \omega +\omega^2 - \mu^2 \bigg),\\
\delta&=\lambda.
\end{align}
We know that the $tanh^{-1}(u)$ terms have no effect on the singularity structure of Eq. (\ref{inceleradial}), and it still has the singularity structure of the DCH equation (\ref{heund}). However, if $c_2 \ne 0$, i.e. in the presence of an EM potential, the form of the radial equation with $F_2(u,c_2)$ does not meet with the DCH equation. Moreover, if we assume that Eq. (\ref{inceleradial}) is a DCH equation, then its parameter $\alpha$ should be zero, therefore it should be possible to transform it to the Mathieu equation using the method we studied in the Appendix. However, when we compare Eq. (\ref{mathieustart}) with the radial equation given in (\ref{yardimci}), we observe that the coefficient of $sinh(2x)$ in Eq. (\ref{yardimci}) has a radial dependence, unlike Eq. (\ref{mathieustart}) which prevents us from using this method to obtain a Mathieu-type solution, or a DCH solution with $\alpha=0$ correspondingly.

We can study Eq. (\ref{inceleradial}) numerically to see the behavior of the radial solutions. We employed the SageMath software system in our numerical analysis \cite{sagemath} as explained in \cite{Birkandan:2017tlh}. As an alternative, one can use the integral series approach given in \cite{Giscard:2020iqg,numericalheun}. It should be noted that the spikes at the far right of some graphs are the artifacts of the numerical solver and they should be disregarded.
\begin{figure}[H]
	\centering
	\includegraphics[scale=0.8]{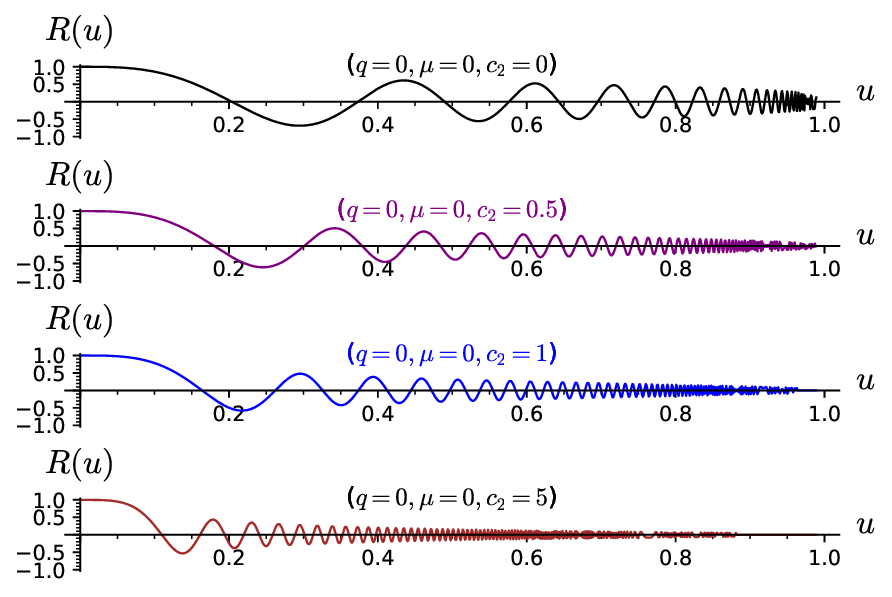} 
	\caption{Numerical solutions of the radial equation (\ref{inceleradial}) with \{$q=0,\mu=0$\} for different $c_2$ values. The numerical values of the other parameters are \{$\lambda=1,a=3,m_1=0.3,m_2=0.5,\omega=10$\}.}
	\label{fig:qmuzero}
\end{figure}
\begin{figure}[H]
	\centering
	\includegraphics[scale=0.8]{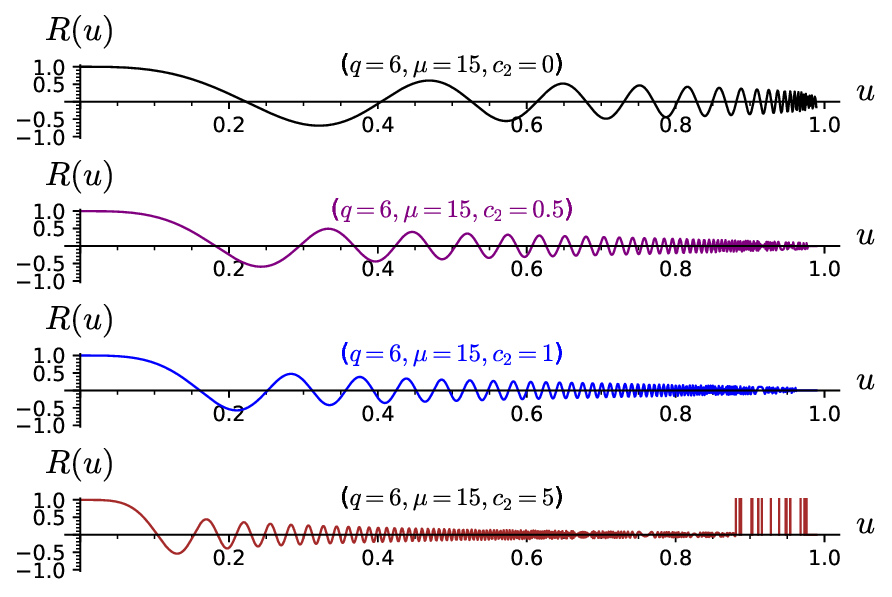} 
	\caption{Numerical solutions of the radial equation (\ref{inceleradial}) with \{$q=6,\mu=15$\} for different $c_2$ values. The numerical values of the other parameters are \{$\lambda=1,a=3,m_1=0.3,m_2=0.5,\omega=10$\}.}
	\label{fig:qmunonzero}
\end{figure}
\begin{figure}[H]
	\begin{subfigure}{.5\textwidth}
		\centering
		% include first image
		\includegraphics[width=1.0\linewidth]{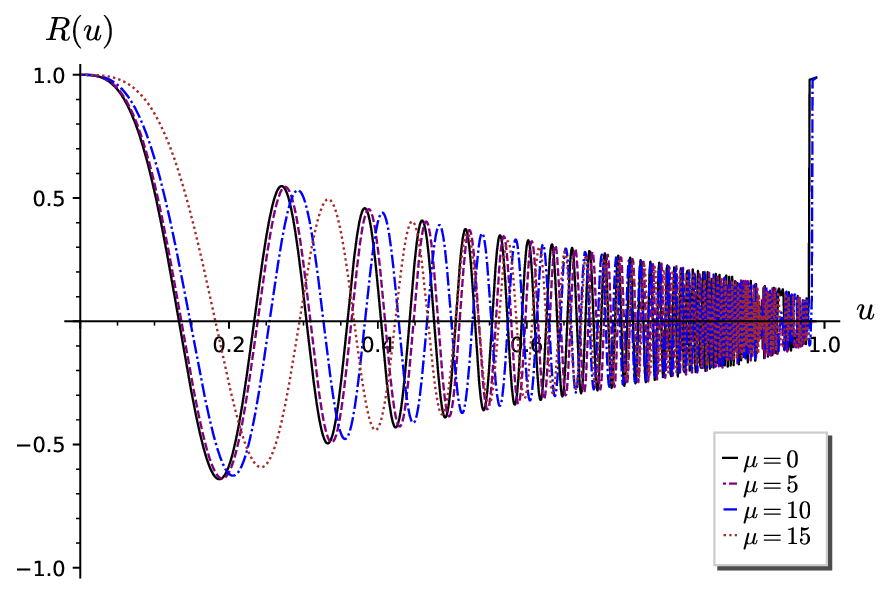}
		\caption{Varying mass ($q=6$).}
		\label{fig:sub-first0}
	\end{subfigure}
	\begin{subfigure}{.5\textwidth}
		\centering
		% include second image
		\includegraphics[width=1.0\linewidth]{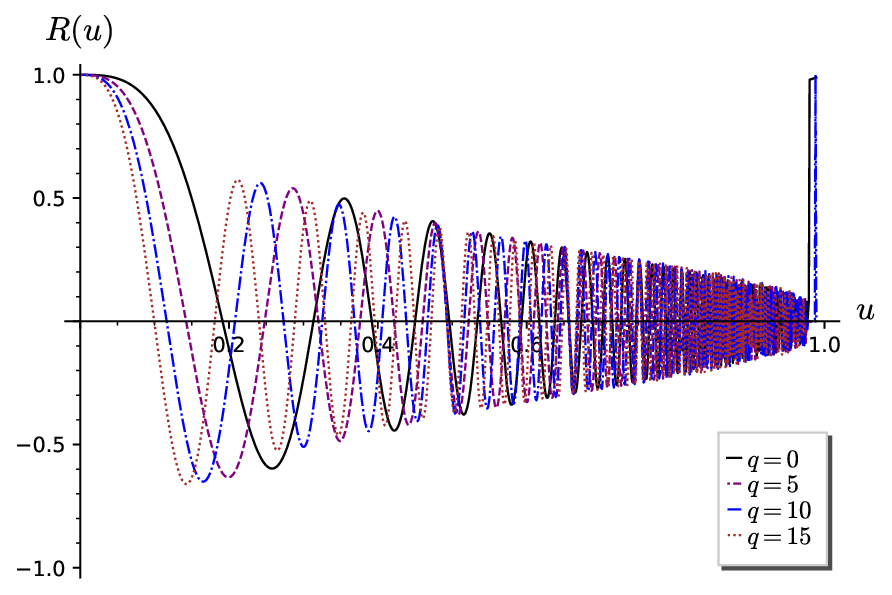}
		\caption{Varying charge ($\mu=15$).}
		\label{fig:sub-second0}
	\end{subfigure}
	\caption{Numerical solutions of the radial equation (\ref{inceleradial}) for several mass and charge values. The numerical values of the other parameters are \{$\lambda=1,a=3,m_1=0.3,m_2=0.5,\omega=10, c_2=0.5$\}.}
	\label{fig:varmuq}
\end{figure}
As one can see from the Figures (\ref{fig:qmuzero}, \ref{fig:qmunonzero}, \ref{fig:varmuq}), changes in the mass and charge values affect the behavior of the solution slightly, and the EM field (via the parameter $c_2$) does not change the general character of the solution except the decaying behavior and the ``effective" radial range where the radial solution has a visible oscillatory nature. We can claim that $c_2$ puts an ``effective" cut-off on the physical range as the solutions approach to zero after a radial distance from the origin.

In Figures (\ref{fig:qmuzero}) and (\ref{fig:qmunonzero}), one can observe that the behavior of the solutions with a non-zero $c_2$ is very similar to the ones with the solutions of the DCH type (with zero $c_2$). Thus we can claim that the solutions with non-zero $c_2$ values might correspond to a form of a ``deformed" DCH equation. Then, we might attempt to find a quasi-exact solution using some approximations.

\begin{figure}[H]
	\begin{subfigure}{.5\textwidth}
		\centering
		% include first image
		\includegraphics[width=1\linewidth]{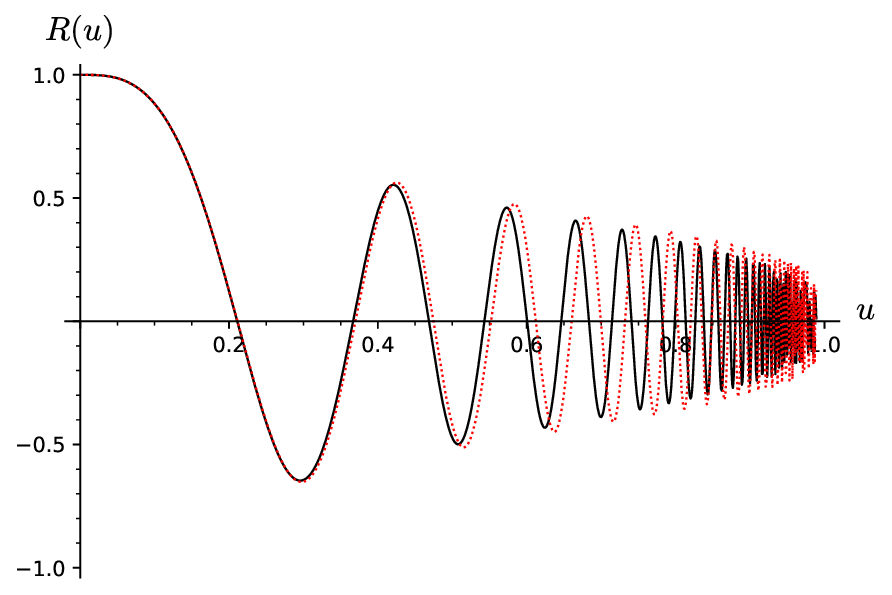}
		\caption{$c_2=0.1$}
		\label{fig:sub-first}
	\end{subfigure}
	\begin{subfigure}{.5\textwidth}
		\centering
		% include second image
		\includegraphics[width=1\linewidth]{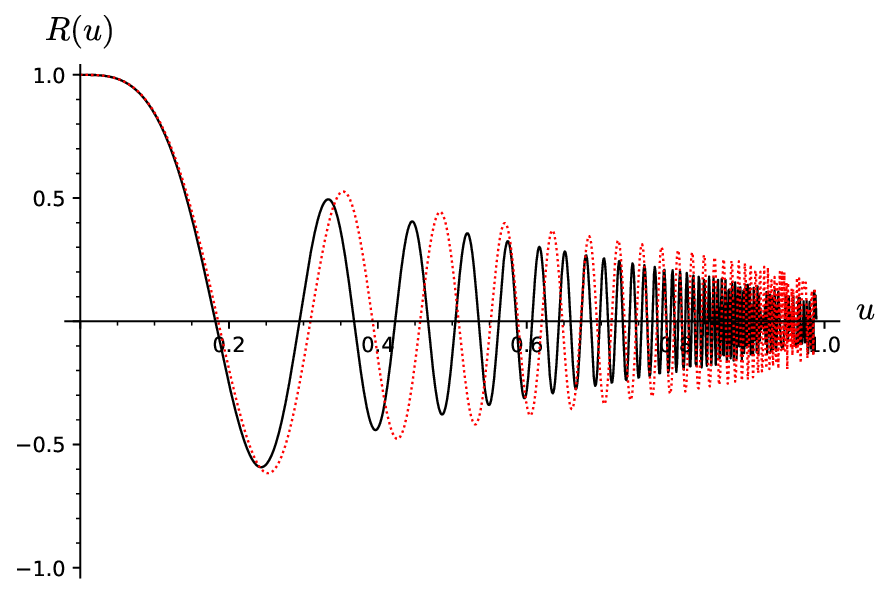}
		\caption{$c_2=0.5$}
		\label{fig:sub-second}
	\end{subfigure}
		\begin{subfigure}{.5\textwidth}
		\centering
		% include third image
		\includegraphics[width=1\linewidth]{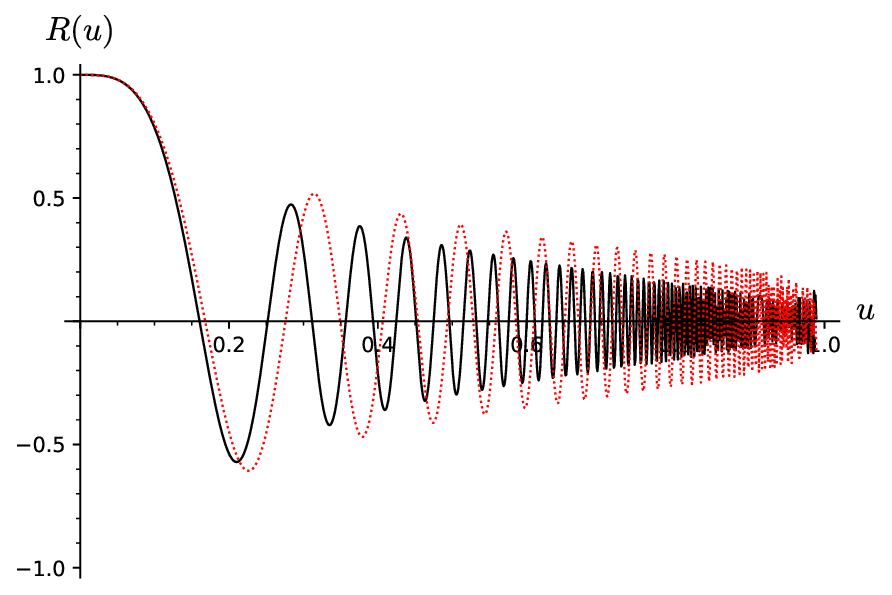}
		\caption{$c_2=1$}
		\label{fig:sub-third}
	\end{subfigure}
		\begin{subfigure}{.5\textwidth}
		\centering
		% include fourth image
		\includegraphics[width=1\linewidth]{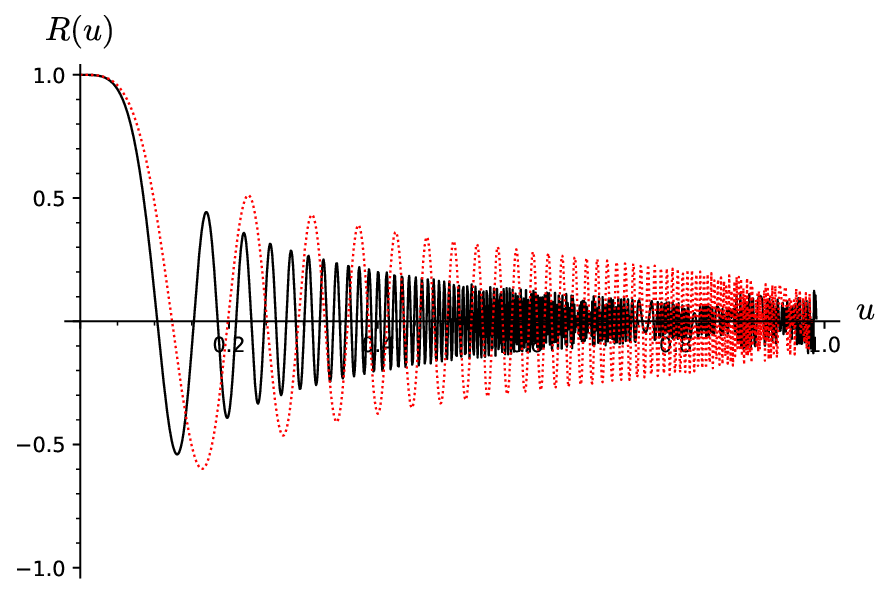}
		\caption{$c_2=5$}
		\label{fig:sub-fourth}
	\end{subfigure}
	\caption{Full radial equation (\ref{inceleradial}) (solid curve) vs. small $u$ approximation determined by (\ref{bas}-\ref{son}) (dotted curve). The numerical values of the other parameters are \{$\lambda=1, a=3, m_1=0.3, m_2=0.5, \omega=10, q=6, \mu=15$\}.}
	\label{fig:fig4}
\end{figure}

For $|u|<1$, we have $tanh^{-1}(u)=\sum_{k=0}^{\infty} \frac{u^{2k+1}}{2k+1}$. Let us make the approximation $tanh^{-1}(u) \sim u$ in the vicinity of $u=0$ and keep only the quadratic term in $u$ in eq. (\ref{F2uc2}) in order to preserve the DCH form. Then we can write this equation as a DCH equation with the parameters
\begin{align}
	\alpha&=0, \label{bas}\\
	\beta&=-a^3 \bigg(\frac{3}{2} q^2 + 2 \sqrt{6} q \omega+  3\omega^2 -\mu^2 \bigg) c_2+ (m_1^2+m_2^2)a^2-\lambda, \label{cicibeta}\\
	\gamma&=- a^2 \bigg(\frac{3}{2} q^2+\sqrt{6} q \omega +\omega^2 - \mu^2 \bigg),\\
	\delta&=\lambda. \label{son}
\end{align}
It should be emphasized that the range cut-off cannot be read from this solution, however, it can serve as a ``quasi-exact" solution for small $u$ values as can be seen from Figure (\ref{fig:fig4}). It can also be seen that our quasi-exact DCH solution requires small $c_2$ values for a better fit.

In this limit, the effect of the EM field can be eliminated when the coefficient of $c_2$ in eq. (\ref{cicibeta}), namely $\frac{3}{2} q^2 + 2 \sqrt{6} q \omega+  3\omega^2 -\mu^2$ is zero.
%%%%%%%%%%%%%%%%%%%%%%%%%%%%%%%%%%%%%%%%%%%%%%%%%%%%%%%%%%%%%%%%%%%%%%%%%%%%%%%%%%%%%
\section{Conclusion}
We studied a charged and massive scalar field in the background of the Nutku-Ghezelbash-Kumar metric which was obtained by the addition of a time coordinate to the Nutku helicoid metric in a non-trivial way, using the source-free Einstein-Maxwell equations.

The charged and massive scalar field could be separated into its radial and angular parts. The angular part was free of the physical parameters charge and mass, and it could be written as a double confluent Heun equation which could also be transformed into a Mathieu equation. However, the full radial part could not be solved in terms of a known function in its general form. 

When the charge, mass, and the parameter associated with the electromagnetic field were set to zero, we recovered the double confluent Heun solution in the literature written for the Nutku metric with a trivially added time coordinate. The charged and massive scalar field without the electromagnetic field parameter also yielded the double confluent Heun equation.

In the presence of the electromagnetic interaction, we could not solve our radial equation in terms of double confluent Heun functions. This was the first case without an exact double confluent Heun solution in the literature based on the Nutku metric.

The full radial equation could be analyzed numerically to see the effect of the physical parameters on the solution. The changes in the charge and mass values affected the solution slightly but more importantly, the change in the electromagnetic field parameter set a cut-off on the effective range of the radial coordinate.

We managed to find a ``quasi-exact" solution for the radial equation when we approximated the arctangent function for small radial values. Neglecting the terms higher than quadratic in radial coordinate, we could write the equation in terms of the double confluent Heun parameters. We also required a small electromagnetic field parameter in order to prevent the effect of the higher-order terms. In this limit, we also found an expression that eliminated the electromagnetic coupling for the special values of the physical parameters.

As they all have $\alpha=0$ in their parameter sets, all equations that could be written in a double confluent Heun form could also be written in terms of a Mathieu equation in our study.
%%%%%%%%%%%%%%%%%%%%%%%%%%%%%%%%%%%%%%%%%%%%%%%%%%%%%%%%%%%%%%%%%%%%%%%%%%%%%%%%%%%%%
\section*{Acknowledgement} 
This paper is dedicated to Prof. Mahmut Horta\c{c}su on the occasion of his 80th birthday.
\newline
%%%%%%%%%%%%%%%%%%%%%%%%%%%%%%%%%%%%%%%%%%%%%%%%%%%%%%%%%%%%%%%%%%%%%%%%%%%%%%%%%%%%%
\section*{Appendix: Transformation of the double confluent Heun equation to the Mathieu equation for $\alpha=0$}
In \cite{Birkandan:2006ac}, the transformation of the DCH equation to the Mathieu equation was studied using the parameters related with the spacetime and the fields. Here, we will give a more general approach to the method by using the general form of the DCH equation with $\alpha=0$ which is associated with all cases in the literature where the Nutku helicoid is involved. It should be noted that the variable names $y$, $z$, and $u$ are not related with the metric coordinates used in the main text.

We will start with the general form of the DCH equation given in Eq. (\ref{heund}) and set $\alpha=0$ as our results and some other results from the literature have this property \cite{Birkandan:2006ac,Birkandan:2007ey,Birkandan:2007cw}. Then the equation becomes
\begin{equation}
	\frac{d^2 H_D}{dx^2}
	+{\frac {2\,{x}^{5} -4\,{x}^{3}+2\,x}{ \left( x-
			1 \right) ^{3} \left( x+1 \right) ^{3}}}
	\frac{dH_D}{dx}
	+{\frac {\beta{x}^{2}+ \gamma x+\delta}{
			\left( x-1 \right) ^{3} \left( x+1 \right) ^{3}}}
	H_D=0.
\end{equation}
Following the procedure given in \cite{Birkandan:2006ac}, we perform a change of variable $x=tanhy$ to get
\begin{equation}\label{mathieustart}
\frac{d^2 H_D}{dy^2}+\bigg(\frac{\beta-\delta}{2}-\frac{\gamma}{2}sinh(2y)-\frac{\beta+\delta}{2}cosh(2y)\bigg)H_D=0,
\end{equation} 
and we set
\begin{align}
A&=-\frac{\gamma}{2},\\
B&=\frac{\beta-\delta}{2},\\
C&=-\frac{\beta+\delta}{2}.
\end{align}
Then we use the transformations $z=e^{-2y}$, $u=\sqrt{\frac{C-A}{C+A}}z$, $w=\frac{1}{2}\big(u+\frac{1}{u}\big)$ consecutively to obtain
\begin{equation}
	(1-w^2)\frac{d^2 H_D}{dw^2}-w\frac{d H_D}{dw}-\bigg(\frac{\beta-\delta}{8}-\frac{\sqrt{(\beta+\delta)^2-\gamma^2}}{8}w \bigg)H_D=0,
\end{equation}
which corresponds to an algebraic form of the Mathieu equation. If we perform another transformation $p=cos^{-1} \bigg(\sqrt{\frac{w+1}{2}} \bigg)$, we obtain
\begin{equation}\label{cicimathieu}
\frac{d^2 H_D}{dp^2}+\bigg( \frac{\delta-\beta}{2} +\frac{\sqrt{(\beta+\delta)^2-\gamma^2}}{2}cos(2p)\bigg)H_D=0,
\end{equation}
which is the standard form of the Mathieu equation \cite{nist}.
%In \cite{Ghezelbash:2017bjs}, the metric function $H(r)$ was found by solving the equation,
%\begin{equation}
%(r^2+a^2)\frac{d^2H(r)}{dr^2}+r\frac{dH(r)}{dr}=0.
%\end{equation}
%After applying the transformations we used for the radial part of the Klein-Gordon equation, namely, $r = asinh(x)$ and $x = tanh^{-1}(u)$, we obtain
%\begin{equation}
%(u^2-1)\frac{d^2H(u)}{du^2}+2u\frac{dH(u)}{du}=0.
%\end{equation}
%%%%%%%%%%%%%%%%%%%%%%%%%%%%%%%%%%%%%%%%%%%%%%%%%%%%%%%%%%%%%%%%%%%%%%%%%%%%%%%%%%%%%

\end{document}